# Chemical Doping and Electron-Hole Conduction Asymmetry in Graphene Devices

*Damon B. Farmer[1,≅], Roksana Golizadeh-Mojarad[2,≅], Vasili Perebeinos[1], Yu-Ming Lin[1], George S. Tulevski[1], James C. Tsang[1], and Phaedon Avouris[1,*]*

[1]IBM T.J. Watson Research Center, Yorktown Heights, NY 10598, USA.

[2]School of Electrical and Computer Engineering, Purdue University, West Lafayette, IN 47906, USA.

[≅] These authors contributed equally to this work.

[*] avouris@us.ibm.com, 914-945-2722

We investigate polyethylene imine and diazonium salts as stable, complementary dopants on graphene. Transport in graphene devices doped with these molecules exhibits asymmetry in electron and hole conductance. The conductance of one carrier is preserved, while the conductance of the other carrier decreases. Simulations based on nonequilibrium Green's function formalism suggest that the origin of this asymmetry is imbalanced carrier injection from the graphene electrodes caused by misalignment of the electrode and channel neutrality points.

**KEYWORDS:** graphene, chemical doping, nonequilibrium Green's function formalism, conductance asymmetry, carrier injection

There has been much progress in the fabrication and understanding of graphene devices.[1,2,3,4,5,6,7,8] However, many key issues still need to be addressed in order to fully exploit the high mobility exhibited by graphene in device technology. Chemical doping is one such subject that requires attention. Several chemical species are known to produce doping effects in mechanically exfoliated graphene.[9,10] These effects typically include the suppression of both electron and hole conduction. Here, we show that this type of doping-induced response is not universal, and in doing so, develop a model to explain the general presence of conduction asymmetry in graphene devices. Polyethylene imine (PEI) and diazonium salts are used as complementary molecular dopants on graphene. We find that these compounds produce doping effects in which the conductance of either electrons or holes is suppressed, but not both. Furthermore, the carrier type that is suppressed is found to have the opposite polarity of the dopant. Our simulations suggest that this conductance asymmetry is caused by imbalanced carrier injection from the device electrodes. We also find that the nature of the doping-induced potential in the graphene channel is critical in determining the type of asymmetry exhibited by the device.

Graphene devices are fabricated using the conventional mechanical exfoliation process to isolate graphene flakes from highly oriented pyrolytic graphite (HOPG).[1] These flakes are deposited on a heavily p-doped Si substrate that is terminated with 300 nm of $SiO_2$. The Si substrate is used as the gate electrode, and the oxide serves as the gate dielectric. Source and drain electrodes are defined on the graphene flakes using electron beam lithography, and deposited onto the substrate by electron beam evaporation. The electrodes consist of a 0.5 nm Ti adhesion layer, followed by a 20 nm Pd layer and a 30 nm Au capping layer. Electrical measurements of these back-gated devices are made at 300 K under a vacuum pressure of 3 x $10^{-7}$ Torr.

Diazonium salts are organic compounds that have previously been employed to separate carbon nanotubes according to their electronic structure via selective covalent attachment.[11] Graphene devices were exposed to a 1 mM solution of 4-bromobenzenediazonium tetrafluoroborate in a 1:1 water/methanol mixture. This was done at 300 K under atmospheric pressure for 2 hours before rinsing with copious amounts of water and methanol. Figure 1a shows the conductance profile of a graphene device before and after exposure to this diazonium salt. The shift of the $G_{ds}$-$V_g$ curve to more positive gate voltages after diazonium exposure is an indication of p-type doping. It can be seen that this doping does not significantly alter the conductance at the neutrality point (minimum conductance point, $V_{NP}$), or drastically change the current-modulating behavior of the device. The lack of conductance suppression suggests an absence of appreciable *sp³* hybridization of the graphene surface, which is in stark contrast to the covalent attachment observed on carbon nanotubes.[12] This may be because the curvature-induced, strained configuration of the nanotube better facilitates covalent reaction with the diazonium cation.[13]

While the diazonium interaction with graphene does not appear be covalent in nature, spectroscopic analysis suggests that it is stronger than simple van der Waals-mediated physical adsorption. Figure 1b compares Raman spectra of a graphene flake before and after diazonium exposure. The spectrum before exposure is indicative of single-layer graphene, exhibiting a 2D vibrational mode intensity (~ 2700 cm$^{-1}$) that is greater than the G vibrational mode intensity (~ 1582 cm$^{-1}$).[14] A disorder related peak is also present at the D vibrational mode (~ 1350 cm$^{-1}$), which is typically observed after the device fabrication process. Multiple peaks with varying intensities around the D mode appear after diazonium functionalization. These peaks indicate the presence of the diazonium compound, and have previously been observed on functionalized glassy carbon and carbon nanotubes.[11,15] In contrast, Raman spectra of the neighboring SiO$_2$ substrate are devoid of this structure, signifying that diazonium selectively adsorbs on the graphene surface. The ratio of the 2D intensity and G intensity ($I_{2D}/I_G$) decreases after diazonium exposure. This is an indication of doping, and is in agreement with the induced p-type behavior observed in our electrical measurements.[16] The selective, p-type adsorption exhibited here suggests that the diazonium binding mechanism is the first step of a two-step mechanism proposed for diazonium functionalization of carbon nanotubes.[17] In this mechanism, the diazonium cation first noncovalently binds to the nanotube surface via partial charge-transfer before covalent attachment is achieved. As mentioned above, the lack of curvature may prevent covalent attachment to graphene. Therefore, diazonium will be left in the intermediate state, doping graphene through charge transfer but not covalently modifying it, in agreement with our observations. Such an interaction does not exist between diazonium and $SiO_2$, which explains why diazonium is only detected on graphene.

Normalizing the $G_{ds}$-$V_g$ curves with respect to $V_{NP}$ reveals a distinct asymmetry in electron and hole conductance (Fig. 1c). Diazonium exposure does not significantly alter hole conduction ($V_g < V_{NP}$), but there is a clear suppression of electron conduction ($V_g > V_{NP}$). This type of doping-induced conductance asymmetry, where conductance of only one carrier type is suppressed, has not been reported previously. Similar investigation of a complementary molecular dopant will therefore help to establish the general pervasiveness of this effect, hence our impetus for studying PEI.

PEI is an amine-rich, electron-donating macromolecule that is known to be an effective n-dopant on carbon nanotubes.[18] Graphene devices were soaked for three hours in a 300 K ethanol solution containing a 20 wt% quantity of PEI. The devices were then rinsed in ethanol to remove excess PEI. Indeed, we find that adsorption of PEI on graphene results in n-type behavior. This is illustrated in Figure 2a, where the $G_{ds}$-$V_g$ curve after PEI exposure exhibits a shift of the neutrality point to more negative gate voltages. In addition, the type of doping asymmetry exhibited after PEI treatment is similar to what was observed after diazonium treatment. In this case, however, the hole conductance ($V_g < V_{NP}$) is suppressed and the electron conductance ($V_g > V_{NP}$) is preserved (Fig. 2b). It is now clear that this asymmetric doping effect is not exclusive to one dopant, but rather the result of a more general transport phenomenon. This is further evidenced by the fact that, in both cases, the charge of the dopant determines the direction of the asymmetry, i.e. whether electron or hole conductance is suppressed.

To further understand this behavior, simulations based on nonequilibrium Green's function formalism (NEGF) are performed to explore the device conditions needed to reproduce the observed conductance asymmetry.[19] In our model, the device is divided into two distinct regions: graphene electrodes underneath metal contacts, and the graphene channel that is exposed to the dopants. We obtain the channel conductance ($G_C$) at the Fermi level ($E_F$) using, $G_C(E_F) = I(E_F)/V$, where,

$$I(E_F) = (e/h) \int_{-\infty}^{+\infty} T(E)(f_L[E-(E_F + eV/2)] - f_R[E-(E_F - eV/2)])dE, \qquad (1)$$

and V is the applied bias across the channel. The Fermi function, $f_{L,R}$, represents the carrier distribution in the left and right contacts, and the transmission coefficient is defined as, $T(E) = trace(\Gamma_L G \Gamma_R G^+)$. In these calculations, the electrochemical potential of the graphene electrodes are pinned to the metal contacts, and the applied gate voltage ($V_g$) changes the Fermi level in the channel via field effect modulation, $E_F = \hbar v_F (\pi C_g V_g)^{1/2}$, where $C_g$ = 115 AF/µm². The Green's function (G) at each energy (E) is found by solving, $G = (EI - H_o - U - \Sigma_L - \Sigma_R)^{-1}$, where $H_o$ is the channel Hamiltonian derived using a π-orbital nearest neighbor tight binding model, U is the doping-induced potential profile in the graphene channel, and I is the unitary matrix. The self energy, $\Sigma_{L,R}$, represents interaction of the semi-infinite graphene electrodes with the channel, and has the general form, $\Sigma_{L,R} = \tau_{L,R} g_{L,R} \tau_{L,R}^+$, where τ is the coupling between the channel and the contacts, and g is the surface Green's function for the electrodes. Metal-induced broadening (η) of the density of states (DOS) is assumed to occur in the graphene electrodes.[20] As a result, g is obtained from the Hamiltonian of the isolated graphene electrode ($H_{electrode}$) using $g = [(E+i\eta)I - H_{electrode}]$. This is evaluated using a recursive Sancho-Rubio method by exploiting the tri-diagonal nature of $H_{electrode}$.[21]

In the simulations, the potential energy of the channel is modified relative to the potential energy of the graphene electrodes in order to simulate the introduction of dopants. We investigate three types of dopant-induced potential barriers: a homogeneous barrier, a barrier produced by short-range scatterers, and a barrier produced by long-range scatterers. The homogeneous barrier is characterized as a constant potential along the length and width of the graphene channel,

$$U(i,j) = U_B \delta_{ij}, \qquad (2)$$

where i and j are lattice site indices. In contrast, short-range scatterers cause the potential barrier to fluctuate across the channel,

$$U(i,j) = [U_B + u(i,j)]\delta_{ij}, \qquad (3)$$

where u(i,j) is the fluctuation parameter. Lastly, long-range scatterers produce a Coulombic potential barrier of the form,

$$U(i,j) = U_B \left\| \sum_{x,y} u(x,y) \left[ \frac{1}{|r(x,y,h) - r(i,j,0)|} \right] \right\|, \tag{4}$$

where u(x,y) is the randomized strength of the scatterers, r is the distance between the scatterers and the graphene lattice sites, and h is the height of the scatterers above the graphene surface.

As illustrated in Figure 3a, transport through the short-range potential results in suppression of both electron and hole conductance. This is due to the multiple reflections (scattering events) experienced by the carriers as they travel through the spatially fluctuating potential. A closer inspection, however, reveals that one branch is suppressed more than the other, depending on the barrier direction. Hole conductance is larger for p-type doping, while electron conductance is larger for n-type doping. By contrast, the homogeneous potential eliminates scattering in the channel altogether, and transport becomes solely dependent on the nature of the electrode/channel interface. As illustrated in Figure 3b, this results in an asymmetric conductance response in which the conductance of only one carrier type is suppressed. Here, p-type (n-type) doping produces a potential barrier that suppresses electron (hole) conductance and preserves hole (electron) conductance. A similar result is obtained for the more realistic case of the long-range scattering potential (Fig. 3c). In this case, however, the potentials of individual scatterers combine to form a channel potential with a high degree of homogeneity, thereby minimizing the amount of scattering in the channel. Regardless of scattering, conductance asymmetry is prevalent in all three cases. Our simulations suggest that the origin of this asymmetry is imbalanced electron-hole injection from the graphene electrodes caused by the doping-induced neutrality point misalignment, $U_B$. This misalignment can be caused by doping in the channel as described above, or alternatively by doping in the electrodes, which is predicted to occur in graphene devices.[22] With respect to the neutrality point of the channel, the presence of $U_B$ creates an unequal density of electrons and holes in the graphene electrodes (Fig. 3 schematics). This results in the asymmetric injection of carriers from the electrodes into the corresponding electron and hole states of the channel.

To further clarify this, two additional types of electrodes are simulated: metal electrodes with a constant DOS ($g = -i\gamma$) and unperturbed, intrinsic graphene electrodes ($\eta = 0$). This allows us to separate the effects caused by the potential $U_B$ from the effects caused by the electrodes. Carrier injection from metal electrodes is found to be symmetric, independent of the potential profile in the channel (Fig. 4a). This is because electrodes with a constant DOS inject the same electron and hole density into the channel. On the other hand, carrier injection from the intrinsic graphene electrodes reproduces the conductance asymmetry seen in Figure 3 (Fig. 4b). The only difference is an extra minimum in the conductance profile, which is due to the neutrality point of the graphene electrodes. To explain this, consider the channel and electrode regions to be two resistors in series, where the total conductance ($G_{ds}$) is proportional to the DOS in the channel ($D_C(E)$) and the DOS in the electrodes ($D_E(E)$),

$$G_{ds} \propto \frac{D_E(E) D_C(E)}{D_E(E) + D_C(E)} . \quad (5)$$

When $D_E(E)$ is constant, as in the case of metal electrodes, $G_{ds}$ will exhibit one minimum corresponding to the channel neutrality point at $E = U_B$ (Fig. 4a schematic). However, when $D_E(E)$ is non-constant, as in the case of intrinsic electrodes, $G_{ds}$ will have two minima, one at $E = 0$, where $D_E(E)$ is minimized, and the other at $E = U_B$, where $D_C(E)$ is minimized (Fig. 4b schematic). Broadening $D_E(E)$ results in two unequal conductance minima because the minimum value of $D_E(E)$ is larger than the minimum value of $D_C(E)$ (Fig. 3 schematics). This, in conjunction with the quadratic dependence of $V_g$ on $E_F$ causes the simulated $G_{ds}$-$V_g$ curves of Figure 3 to only exhibit one conductance minimum. Since this is in agreement with our experimental observations, broadening the graphene electrode DOS in the simulations of the three doping scenarios is justified.

From this analysis we conclude that the doping-induced conductance asymmetry observed in our experiments is caused by a combination of the neutrality point misalignment at the electrode/channel interface and the non-constant DOS of the graphene electrodes. The simulations predict two different types of conductance asymmetry based on the nature of the induced channel potential and the corresponding presence or absence of scattering.  Homogeneous potentials created by long-range scatterers cause conductance suppression of only one carrier type, while inhomogeneous potentials created by short-range scatterers cause conductance suppression of both carrier types.  According to this, PEI and diazonium salts both behave as long-range scatterers on graphene.  This agrees with the ionic (charge-transfer) nature of dopant adsorption discussed above.  Finally, it is important to reiterate that if metal-induced doping of graphene electrodes occurs, this model predicts conductance asymmetry even when the channel is intrinsic.  This explains recent experimental work that identified the electrode/channel interface as the region responsible for conductance asymmetry in intrinsic devices.[23]

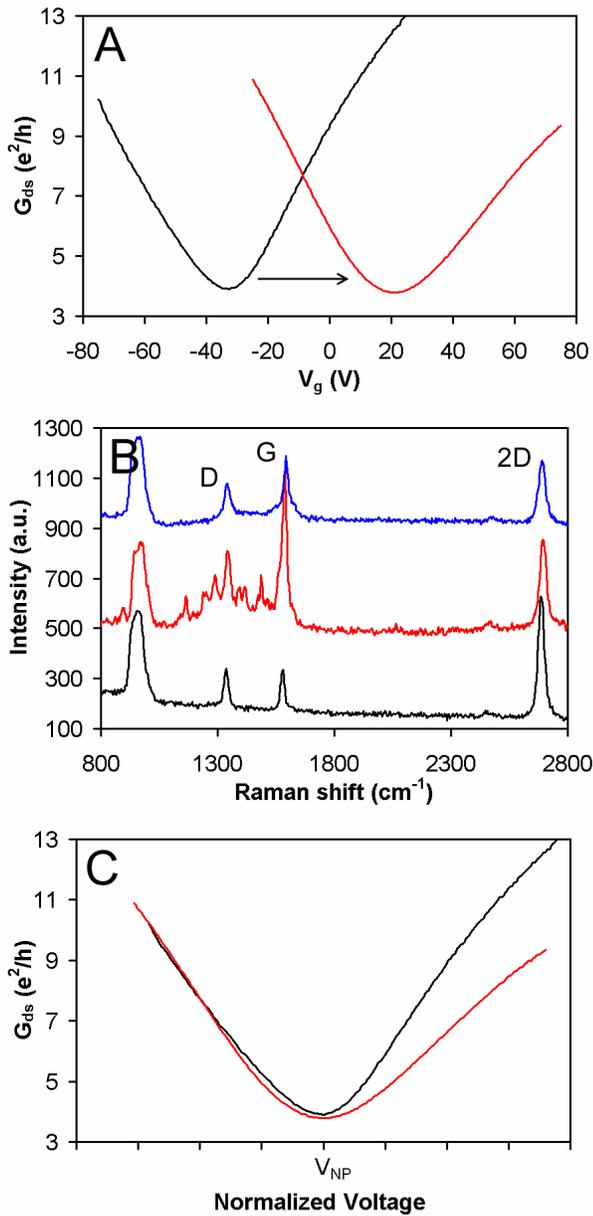

**Figure 1.** The effects of diazonium doping on graphene devices. A) Source-drain conductance vs gate voltage ($G_{ds}$-$V_g$) curves of a graphene flake before (black) and after (red) doping with a diazonium salt. The source-drain bias for these measurements was 10 mV. B) Raman spectra of a graphene flake before (black) and after (red) doping with diazonium. After doping, this flake is annealed at a temperature of 620 K for 2 hours in a 200 mTorr vacuum of flowing Ar (blue). The diazonium peaks disappear after annealing, and the $I_{2D}/I_G$ ratio increases, indicating desorption of the molecules from the graphene surface. C) Normalized $G_{ds}$-$V_g$ curves of the graphene device in (A) showing the dopant-induced

conductance asymmetry in which hole conductance is preserved.  The electrical measurements presented here were made before the 620 K anneal.

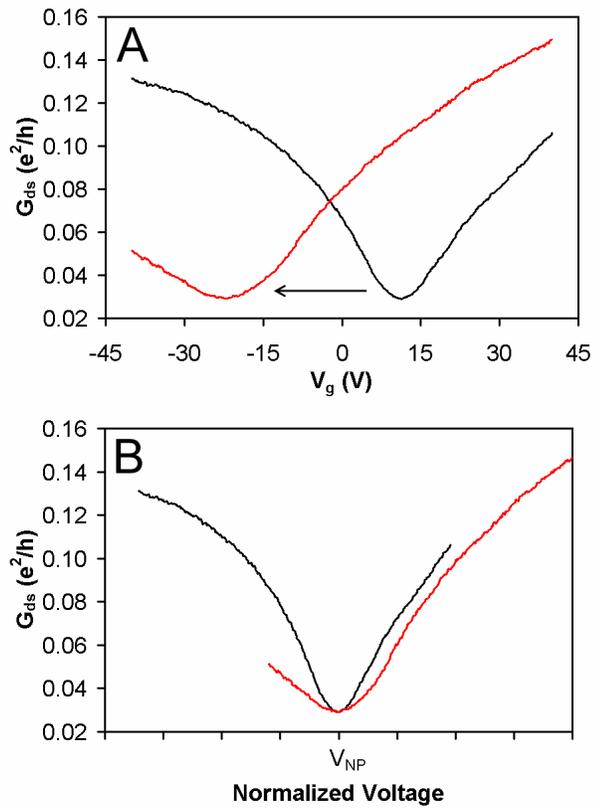

**Figure 2.** The effects of PEI doping on graphene devices. A) $G_{ds}$-$V_g$ curves of a graphene nanoribbon before (black) and after (red) PEI doping. The source-drain bias for these measurements was 10 mV. B) Normalized $G_{ds}$-$V_g$ curves of the graphene device showing the dopant-induced conductance asymmetry in which the electron conductance is preserved.

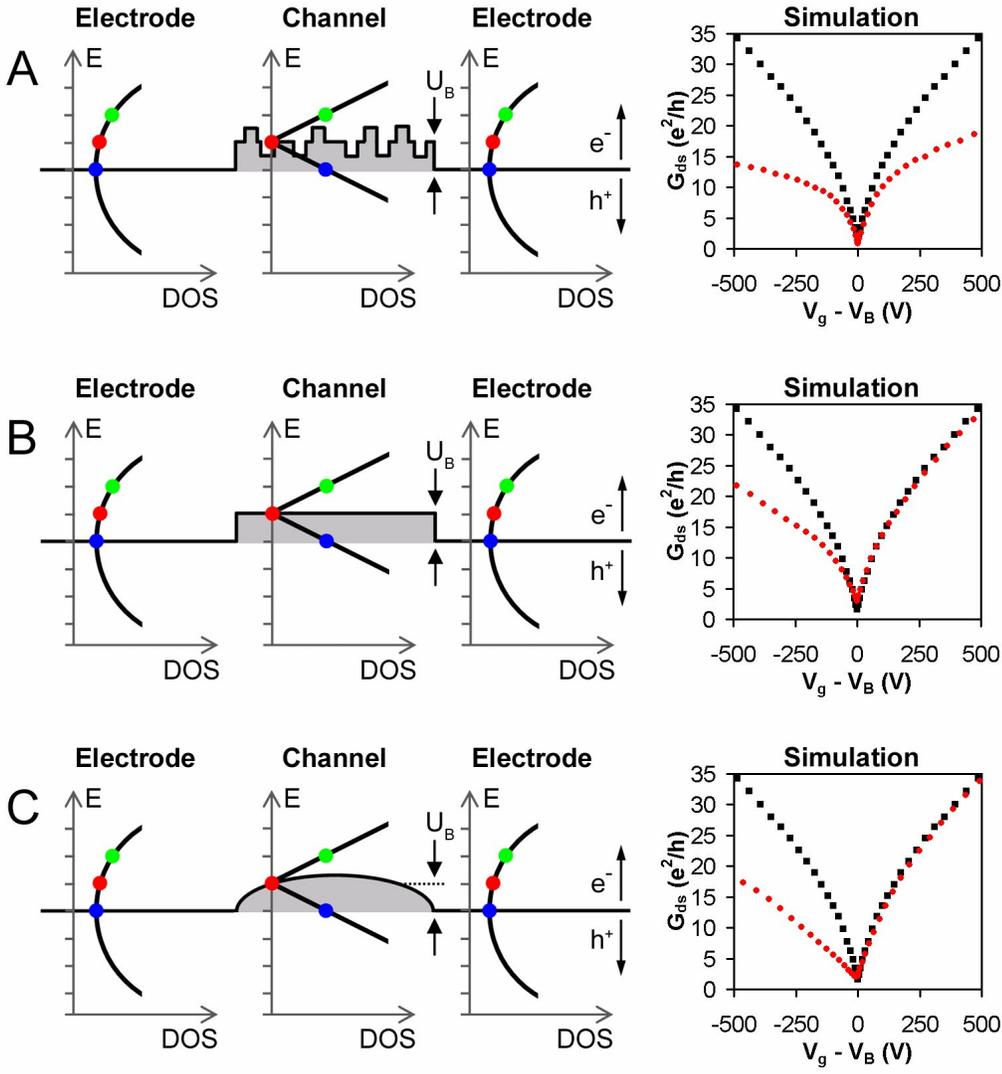

**Figure 3.** NEGF-based simulations comparing three types of graphene channel potentials. The simulated conductances corresponding to each potential (red circles) are compared to the conductance of intrinsic graphene (black squares). A) Inhomogeneous potential caused by short-range scatterers. The fluctuation parameter u(i,j) ranges between $U_B$ and $-U_B$. B) Homogeneous potential. C) Quasi-homogeneous potential caused by long-range scatterers. The scatterer strength u(x,y) ranges between 0 eV-m and 1.2 eV-m, and h = 0.3 nm. In these three simulations, $\eta$ = 0.2 eV, $U_B$ = 0.5 eV, $V_B$ = $U_B^2(\pi \hbar^2 v_F^2 C_g)^{-1}$, and the respective potentials U(i,j) exist on every graphene lattice site. The accompanying schematics show the misalignment of the electrode and channel neutrality points by an amount $U_B$, and broadening of the DOS in the graphene electrodes. The red markers correspond to the neutrality points in the channel. While the blue and green markers represent an equivalent DOS in the channel, the corresponding DOS in the electrodes are different. This results in imbalanced injection of electrons and holes, which leads to asymmetric conductance with respect to the channel neutrality point.

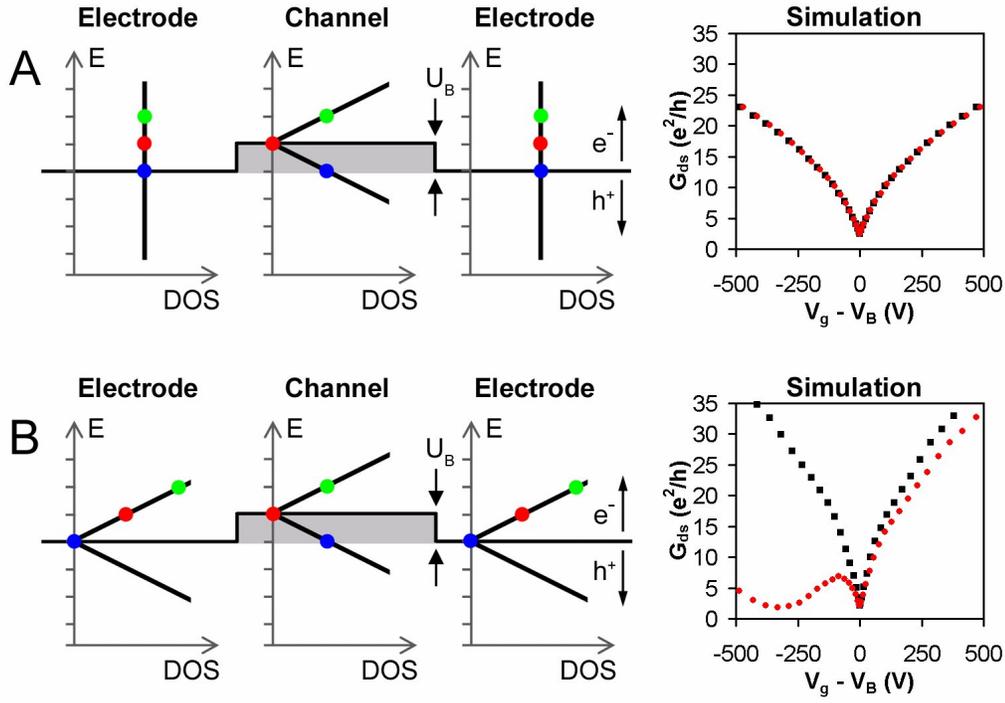

**Figure 4.** NEGF-based simulations showing the effect of the electrode properties on carrier transport. (A) Homogeneous channel potential with metallic electrodes. B) Homogeneous channel potential with intrinsic graphene electrodes. The simulated conductances for these two situations (red circles) are compared to intrinsic graphene (black squares), and the corresponding schematics are also presented. In these simulations, $\gamma = 0.2$ eV for the metallic electrodes, $U_B = 0.5$ eV, and $V_B = U_B^2(\pi \hbar^2 v_F^2 C_g)^{-1}$.

Table of Contents Graphic:

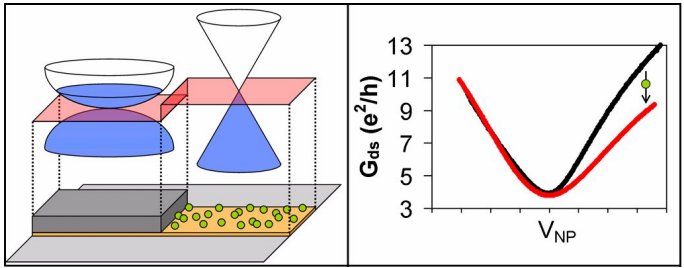